\documentclass[aps,prb,showpacs,superscriptaddress,twocolumn]{revtex4}

\usepackage{graphicx}
\usepackage{longtable}
\usepackage[ansinew]{inputenc}
\usepackage{color}

\newcommand{\diff}{\ensuremath{\text{d}}}
\newcommand{\imp}{\rm{imp}}

\newcommand{\sign}{\,{\rm{sign}\,}}

\newcommand{\beq}{\begin{eqnarray}}
\newcommand{\eeq}{\end{eqnarray}}

\renewcommand{\Re}{{\rm Re}\,}
\renewcommand{\Im}{{\rm Im}\,}

\begin{document}

\title{Theory of Fano Resonances in Graphene: The Kondo effect probed by STM}

\author{T. O. Wehling}
\email[]{twehling@physnet.uni-hamburg.de}
\affiliation{1. Institut f\"{u}r Theoretische Physik I,
Universit\"{a}t Hamburg, D-20355 Hamburg, Germany}
\author{H. P. Dahal}
\affiliation{Theoretical Division, Los Alamos National Laboratory, Los Alamos, New Mexico 87545,USA}
\author{A. I. Lichtenstein}
\affiliation{1. Institut f\"{u}r Theoretische Physik I,
Universit\"{a}t Hamburg, D-20355 Hamburg, Germany}
\author{M. I. Katsnelson}
\affiliation{Institute for Molecules and Materials,
Radboud University Nijmegen, NL-6525 AJ Nijmegen, The Netherlands}
\author{H. Manoharan}
\affiliation{ Department of Physics, Stanford Institute for
Materials and Energy Science, Stanford University, Stanford, CA 94305
USA}
\author{A. V.  Balatsky}
\email[]{avb@lanl.gov, http://theory.lanl.gov}
\affiliation{Theoretical Division, Los Alamos National Laboratory, Los Alamos, New Mexico 87545,USA}
\affiliation{Center for Integrated
Nanotechnologies, Los Alamos National Laboratory, Los Alamos, New
Mexico 87545,USA}

\begin{abstract}
We consider the theory of  Kondo effect and Fano factor energy
dependence for magnetic impurity (Co) on graphene. We have
performed a first principles calculation and find that the two
dimensional $E_1$ representation made of $d_{xz},d_{yz}$ orbitals
is likely to be responsible for the hybridization and ultimately
Kondo screening for cobalt on graphene. There are few high
symmetry sites where magnetic impurity atom can be adsorbed. For
the case of Co atom in the middle of hexagon of carbon lattice we
find anomalously large Fano  $q$-factor, $q\approx 80$ and
strongly suppressed coupling to conduction band. This anomaly is a
striking example of quantum mechanical interference related to the
Berry phase inherent to graphene band structure.
\end{abstract}

\date{\today}

\maketitle Scanning tunneling microscopy (STM) allows us to probe
the electronic properties of conducting materials with atomic
scale spacial resolution. This experimental technique is
particularly well suited to study electron-correlation phenomena.
One of the most famous correlation phenomena is the Kondo effect
arising from a localized magnetic moment being screened by the
conduction electrons \cite{HewsonBook}. It results in a sharp
Abrikosov-Suhl resonance in local density of states (LDOS) of the
impurity at the Fermi level and below a characteristic Kondo
temperature, $T_{\text{K}}$. While the Kondo effect is well
understood for impurities in bulk materials and simple model
systems, STM has been substantial for revealing the intricacies of
Kondo effect at conventional metal surfaces. One well known
example is the Kondo effect caused by Co ad-atoms on a Cu $(111)$
surface \cite{Madhavan98}: It shows that the rich electronic
structure of three-dimensional metals like Cu, in general, makes
impurity effects at their surfaces \cite{Knorr02,LimotCM,Neel}
depending strongly on atomistic details and requires to understand
interaction mechanisms in detail.

Graphene - a monolayer of carbon atoms arranged in a honeycomb
lattice - is the first truly two dimensional material
\cite{Novoselov_science2004} and provides a two-dimensional
electron gas with distinct and highly symmetric low energy
electronic structure: At two non-equivalent corners of the
Brillouin zone, K and K', the linearly dispersing valence and
conduction band touch forming a conical point and leading to the
Berry phase $\pi$ \cite{Geim2005,Zhang2005}. Thus, electronic
excitations in graphene resemble massless Dirac fermions with the
speed of light being replaced by the Fermi velocity $v_{\rm
f}\approx c/300$. Therefore, graphene provides an important model
system for understanding quantum effects in reduced dimensions and
in presence of an ``ultra-relativistic'' conduction electron bath.

A theoretical study showed that even in undoped graphene the Kondo
effect can exist above a certain critical coupling despite the
linearly vanishing density of states
\cite{fradkin-1996,Sengupta07,balseiro-2009}, a situation very
similar to magnetic impurities in the pseudogap phase of high Tc
superconductors \cite{fradkin-1996,RMP_Balatsky}. Moreover,
back-gating\cite{Novoselov_science2004} as well as chemical
doping\cite{SchedinGassensors,NanoLettAds} allows one to control
the chemical potential in graphene and to tune Kondo physics and
electron tunneling in this way.

In this paper, we address how the Kondo effect manifests in STM
experiments on graphene and why Fano resonances in the STM spectra
can depend unusually strong on the chemical potential as well as
the real space position of the impurity. To this end, we firstly
consider the single impurity Anderson model with graphene
providing the host electronic structure and compare to simple
model of usual metal surface. With this background, we turn to a
more realistic ab-initio based description of magnetic impurities
on graphene and discuss the case of Co ad-atoms as in the recent
experiment by Manoharan et al.\cite{Manoharan_Co_Graphene}. By
comparison to Co on Cu (111), an extensively studied
system\cite{Knorr02,LimotCM,Neel} posessing also hexagonal
symmetry of the surface, demonstrate the particular importance of
impurity induced resonances in graphene. Furthermore, we analyze
impurities being bound to different sites of the graphene lattice
and show that there one can expected a strong adsorption site
dependence of Fano factors in STM experiments.

\textit{Model for electron tunneling close to impurities.}  The
$\pi$-band the tight-binding Hamiltonian of graphene reads as
\cite{Wallace-1947}
\begin{equation}
\label{eqn:TB-real-sp} \hat{H_0}=-t\sum_{<i,j>}\left(a^\dagger_i
b_j+b^\dagger_j a_i\right),
\end{equation}
where $a_i$ and $b_i$ are the Fermi operators of electrons in the
carbon  $p_z$ orbital of sublattice atoms A and B in the cell at
$R_i$, respectively. The sum includes all pairs of
nearest-neighbor carbon atoms and $t\approx 2.7$ eV is the hopping
parameter. With the Fourier transformed operators $a_k$ ($b_k$),
defined by
$a_i=\int_{\Omega_B}\frac{\diff^2k}{\Omega_B}e^{ikR_i}a_k$ and
$b_i$ analogously, the Hamiltonian can be rewritten as
\begin{equation}
\hat{H_0}  = \int_{\Omega_B}\frac{\diff^2k}{\Omega_B} \Psi^\dagger_k H_k\Psi_k \text{ with } \Psi(k)=\left[\begin{array}{c}a_k \\
 b_k\end{array}\right],
\end{equation}
 $H_k$ is the $k$-dependent $2\times 2$ matrix
\begin{equation}
\label{eqn:TB_graphene_2}
H_k  =  \left( \begin{array}{cc}
0 & \xi(k)\\
\xi^\ast(k)&  0\end{array} \right),
\label{EQ:Ham0}
 \end{equation}
with $\xi(k)=-t\sum_{j=1}^3e^{ik(b_j-b_1)}$, $b_j$ ($j=1,2,3$) are
the vectors  connecting neighboring atoms \cite{Semenoff-1984},
and $\Omega_B$ is the area of the Brillouin zone. An impurity
contributing a localized orbital,
$\hat{H}_{\imp}=\epsilon_{\imp}\sum_\sigma d_\sigma^\dagger
d_\sigma+U n_\uparrow n_\downarrow$, with Fermi operator $d$,
energy $\epsilon_{\imp}$ and on-site Coulomb repulsion $U$ is
considered. Its hybridizations with the graphene bands is
described by, $\hat{V}=\sum_{k,\sigma} \Psi^\dagger_{k,\sigma} V_k
d_\sigma + \rm{h.c.}$. This problem has been extensively discussed
for normal metals and is usually called ``Anderson impurity
model'' \cite{HewsonBook}.

In this framework, the connection between a tip and a sample in
the STM experiment can be expressed by the transfer Hamiltonian
\begin{equation}
 M = \sum_{\sigma}(M_{dt}d_\sigma^\dagger t_\sigma+{\rm{H.c.}})+\sum_{\nu\sigma}(\Psi_{k\sigma}^\dagger M_{k t}t_\sigma+{\rm{H.c.}})
\end{equation}
describing tunneling of electrons from and to the STM tip with the
tunneling matrix elements $M_{dt}$ and $M_{kt}$ and the Fermi
operators $t_\sigma$ ($t_\sigma^\dagger$) for electrons in the STM
tip.

The Fano $q$-factor in the STM ${\rm d}I/{\rm d}V$ spectra can be
understood in terms of this model by using the equation of motion
approach from Ref. \onlinecite{Madhavan01}. One finds
\begin{equation}
q=\frac{A}{B}
\end{equation}
with
\begin{equation}
A=M_{dt}+\sum_{k}M_{k t}V_{k}{\rm{P}}\left(\frac{1}{E_{\rm
F}-\epsilon_\nu}\right) \label{aq}
\end{equation}
and
\begin{equation}
B=\pi\sum_k M_{k t}V_{k}\delta(E_{\rm F}-\epsilon_\nu),
\label{bq}
\end{equation}
where $\rm{P}$ is the principle value symbol.

 To obtain qualitative insights we proceed by simplifying
these expressions:  For a conduction electron state $k$, denote
its probability density integrated about an atomic sphere centered
at the Co-atom by $|\Psi_k|^2$. Then, if the tip is directly above
the Co impurity, one can approximate
\begin{equation}
 M_{k t}V_{dk}=|\Psi_k|^2MV
\label{eqn:MVapprox}
\end{equation}
with $MV$ independent of $k$. Thus, Eq. (\ref{aq}) and Eq. (\ref{bq}) yield \begin{equation}A=M_{dt}+MV\,\Re\,G(E_{\rm F})\end{equation} and \begin{equation}B=MV\,\Im\,G(E_{\rm F})\end{equation} with the local conduction electron Green function $G(E)=\sum_\nu\frac{|\Psi_k|^2}{E-\epsilon_k-i0^+}$. As argued in Ref \onlinecite{Neel}, this simple model has proved successful to describe Fano factors for CoCu$_n$ clusters on Cu(111) and will be used here to understand Fano resonances in graphene and why they are different to metals like Cu.

\textit{Density of states effects on the Fano factor.} Usually,
the magnetic orbitals of the ad-atom are strongly localized
resulting in $|M_{dt}|\ll|M|$ and consequently
\begin{equation}
 q\approx\,\Re\,G(E_{\rm F})/\Im\,G(E_{\rm F}).
\label{eqn:q-approx-G}
\end{equation}
In a metal with bandwidth D and constant density of states (DOS)
in the vicinity of the impurity, $\Im\,G(\epsilon)=-\pi/2D$ if
$-D<\epsilon<D$, we obtain
$q\approx-\frac{1}{\pi}\ln\left|\frac{D+E_{\rm F}}{D-E_{\rm
F}}\right|\approx-\frac{2E_{\rm F}}{\pi D}$. Hence, $|q|<1$ and
the Kondo effect manifests in STM on normal metals as
anti-resonance close to $E_{\rm F}$ as long as $|M_{dt}|\ll|M|$.
This is very different for graphene:

The graphene DOS is $N^{\rm g}_0(E)=\frac{|E|}{D^2}\cdot\Theta(D-|E|)$ resulting $G(E)=\frac{E}{D^2}\ln\left|\frac{E^2}{D^2-E^2}\right|-i\pi N^{\rm g}_0(E)$ and \begin{equation}
 q\approx-\frac{2\sign(E_F)}{\pi}\ln\left|\frac{E_{\rm F}}{D}\right|.
 \label{eqn:q-graph}
\end{equation}
This results follows directly from linearity of $N(E)$ and the
Kramers-Kronig relations. As $D\approx 6$\,eV and usually $E_{\rm
F}\lesssim 0.5$\,eV, the q-factor can be $q \gg 1$ and the Kondo
effect may manifest in STM as \textit{resonance} instead of an
anti-resonance even for $|M_{dt}|\ll|M|$. This is in contrast to a
normal metal, where predominant tunneling into the conduction
electron states results in a Kondo-antiresonance in STM. Moreover,
Eq. (\ref{eqn:q-graph}) demonstrates that the q-factor in graphene
can be expected to depend strongly on the chemical potential.

\textit{Energy dependence of the asymmetry factor.} Any impurity being coupled to graphene leads to characteristic resonances in the local density of states in the vicinity of the impurity\cite{imp_loc-2007-} and may consequently alter the Green functions to be inserted into Eq. (\ref{eqn:q-approx-G}). To understand the role of resonances in the local electronic structure for the q-factor, we illustrate the situation of a realistic impurity by comparing the experimentally important cases of Co on graphene and Co on Cu (111).

For a realistic description of these systems we performed density
functional calculations within the generalized gradient
approximation (GGA)\cite{Perdew:PW91} on $6\times 6$ graphene
supercells containing one Co ad-atom as well as on Cu(111) slabs
containing $5$ Cu layers and one Co ad-atom. The Vienna Ab Initio
Simulation Package (VASP) \cite{Kresse:PP_VASP} with the projector
augmented wave (PAW) \cite{Bloechl:PAW1994,Kresse:PAW_VASP} basis
sets has been used for solving the resulting Kohn-Sham equations.
In this way we obtained relaxed structures for both systems. In
particular, we found that GGA predicts Co to sit above the middle
of a hexagon on graphene.

To estimate the Fano $q$-factors we extracted the orbitally
resolved Green functions at the impurity site using atomic
orbitals naturally included in the PAW basis sets: The projectors
$\left\langle d_{i}|\psi _{n\mathbf{k}}\right\rangle$ of orbitals
$\left| d_{i}\right\rangle$ localized at the impurity atoms onto
the Bloch eigenstates of the Kohn-Sham problem $\left|\psi
_{n\mathbf{k}}\right\rangle$ are available when using PAW as
implemented in the VASP and these give the local Green functions
according to
\begin{equation}
 G_{ij}(\epsilon)=\sum_{n\mathbf{k}}\frac{\left\langle
d_{i}|\psi _{n\mathbf{k}}\right\rangle \left\langle \psi _{n\mathbf{k}%
}|d_{j}\right\rangle }{\epsilon+i\delta-\varepsilon _{n\mathbf{k}}}. \label{eqn:G0}
\end{equation}
So, we employ here the same representation of localized  orbitals
as used within the LDA+U-scheme implemented in the VASP-code
itself or as discussed in the context of LDA+DMFT in Ref.
\onlinecite{PAW-DMFT}.

\begin{figure}[tb]
 \begin{minipage}{.49\linewidth}
a)
 \includegraphics[width=0.98\linewidth]{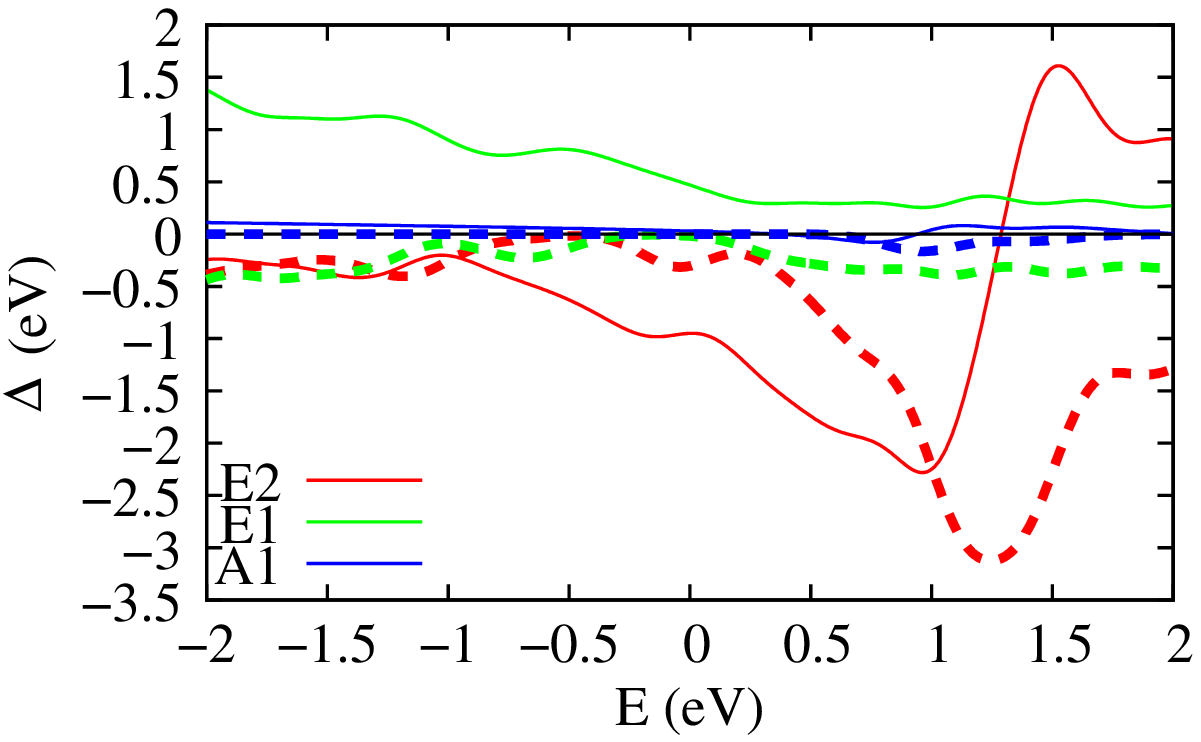}
\end{minipage}
\begin{minipage}{.49\linewidth}
b)
 \includegraphics[width=0.98\linewidth]{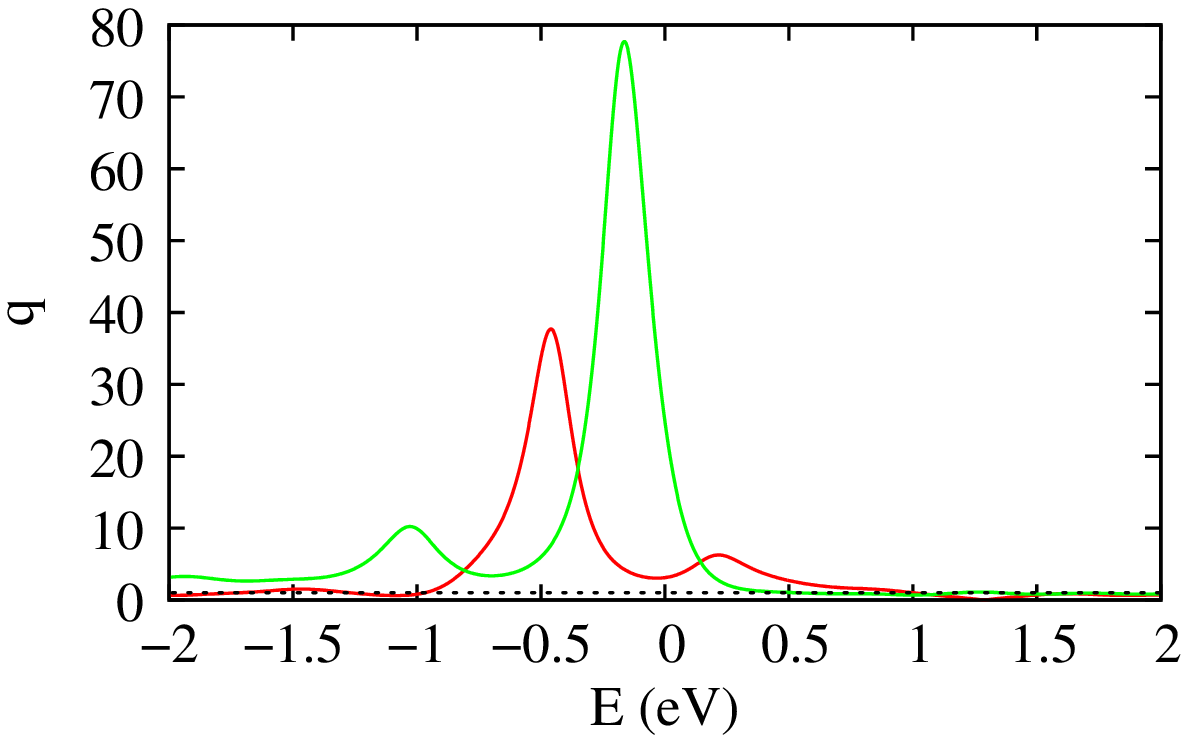}
\end{minipage}

\begin{minipage}{.49\linewidth}
c)
 \includegraphics[width=0.98\linewidth]{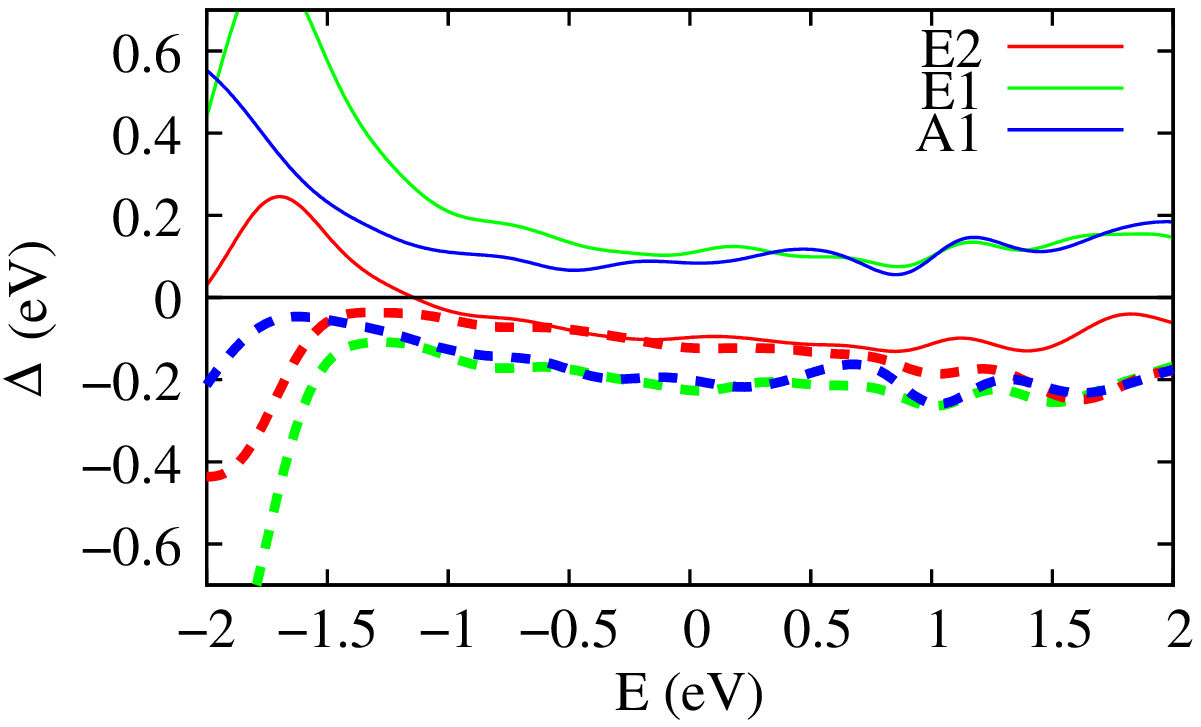}
\end{minipage}
\begin{minipage}{.49\linewidth}
d)
 \includegraphics[width=0.98\linewidth]{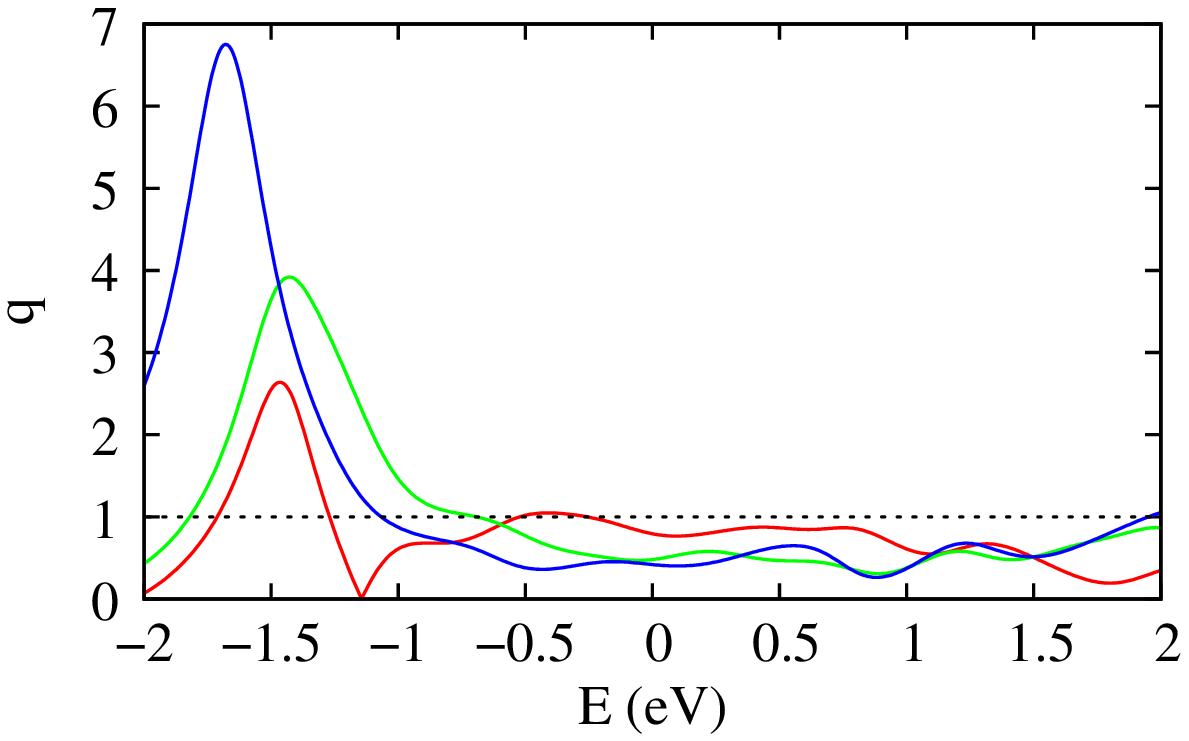}
\end{minipage}
\caption{\label{fig:q-Delta}Hybridization functions $\Delta$ (a,c)
and calculated asymmetry factors $|q|$ for Co on graphene (upper
panel) and Co on Cu (lower panel). For the hybridization functions
$\Re \Delta$ is plotted as solid line, $\Im \Delta$ dashed.}
\end{figure}

The local Green functions at the impurity sites as defined in  Eq.
(\ref{eqn:G0}) are $5\times 5$ matrices which can used to obtain
the hybridization function $\Delta(\epsilon)$ of the impurity:
\begin{equation}
 G^{-1}(\epsilon)=\epsilon+i\delta-\Delta(\epsilon).
\end{equation}
Hence, $\Delta(\epsilon)$ are also $5\times 5$ matrices describing
hybridization of  5 d-electrons of Co. In the particular case of
Co on Cu(111) and graphene, which are both hexagonal surfaces,
these matrices are diagonal and decompose into degenerate blocks
of two 2-dimensional and one 1-dimensional representations,
transforming under the rotation group $C_{6v}$ as E$_1$, E$_2$,
and A$_1$. These components of the hybridization function are
depicted in Fig. \ref{fig:q-Delta} a and c. At energies close to
the Fermi level of graphene all graphene states are in the
vicinity of the two Dirac points. These states transform under
C$_{6v}$ according to E$_1$ and E$_2$. Hence, the hybridization of
the A$_1$ impurity orbital to the graphene bands is strongly
suppressed. Moreover, the crystal field splitting appears to be
such that the E$_1$ orbitals (d$_{xz}$ and d$_{yz}$) are highest
in energy by approx 0.5-1eV as compared to the other d-orbitals.
So, the E$_1$ orbitals are expected to determine the q-factor in
STM experiments probing the Kondo effect of Co on graphene.

For Co on graphene Fig. \ref{fig:q-Delta} shows that $|\Im\Delta(\epsilon)|\ll|\Re\Delta(\epsilon)|$ in the vicinity of $\epsilon=0$, which is the Fermi level for undoped graphene. This is very different from the case of Cu, where $|\Im\Delta(\epsilon)|$ and $|\Re\Delta(\epsilon)|$ are mainly on the same order. Using
\begin{equation}
 \Delta(\epsilon)=\sum_{k}\frac{|V_{k}|^2}{\epsilon+i\delta-\epsilon_k}
\end{equation}
in combination with Eq. (\ref{eqn:MVapprox}) and $|M_{dt}|\ll|M|$ we arrive at
\begin{equation}
 q\approx\,\Re\,\Delta(E_{\rm F})/\Im\,\Delta(E_{\rm F}).
\end{equation}
Within this approximation the projectors and  eigenenergies
obtained from DFT allow for an ab-initio prediction of
$q$-factors. The Fig. \ref{fig:q-Delta} b and d show the
$q$-factors predicted for channels of different C$_6$ rotational
symmetry as calculated for Co on graphene and Cu, respectively, as
function of the resonant energy $E$. $E=0$ corresponds here to the
Fermi level of the undoped system. The calculated $q$-factors for
Co on Cu(111) are typically on the order of $q\lesssim 1$ without
pronounced energy dependence. This is in contrast to graphene,
where $q>1$ in a wide energy rage and $q$ is strongly energy
dependent. So, $q$ is expected to be strongly sensitive to local
changes in the chemical potential of graphene, which can be caused
by gate voltages, chemical doping or substrate effects. We also
point out that this discussion goes well beyond a simple tight
binding and linearized dispersion analysis. Presence of defect can
substantially change local bands and DOS.

\textit{Site dependence of hybridization matrix elements.} For Co
on graphene, we saw that graphene's Fermi  surface being made up
by states transforming as E$_1$ and E$_2$ under $C_6$ lead to
particular Co orbitals being decoupled from the graphene bands.
This special symmetry of graphene's Fermi surface makes the
$q$-factors seen in Kondo resonances in STS particularly dependent
of the precise atomic arrangement of the magnetic impurity. This
can be illustrated by the site dependence of $q$-factors for an
Anderson impurity sitting , on top of C and on a bridge site in
the middle of a hexagon, respectively.

At each adsorption site, a spherically symmetric  s-wave impurity
can be modelled by equal hopping matrix element $V_i$ to all
adjacent sites. For such an Anderson impurity on top of a carbon
atom or at a bridge site we obtain \begin{equation}
 V_k=\left(\begin{array}{c}
 V_i \\ 0
\end{array}
\right) \text{ and } V_k=\left(\begin{array}{c}
 V_i \\ V_i
\end{array}
\right),
\end{equation}
respectively, by translating that nearest neighbor hopping into
the matrix formalism of Eq. (\ref{eqn:TB_graphene_2}) and
performing the Fourier transformation. Combining this with Eq.
(\ref{bq}) results in $B\sim O(E_{\rm F})$ for $E_{\rm
F}\rightarrow 0$. This situation corresponds to Eq.
\ref{eqn:q-graph} with $q$ being enhanced as $E_{\rm F}\rightarrow
0$.

For the impurity in middle of the hexagon the Fourier transformed hopping reads
\begin{equation}
 V_k=\left(\begin{array}{c}
 \xi^\ast(k) \\ \xi(k)
\end{array}
\right).
\end{equation}
As the dispersion does, this coupling vanishes linearly  when
approaching the Brillouin zone corners K and K'. As a consequence,
any possible Kondo resonance due to such an impurity will lead to
$q\gg 1$: Eq. (\ref{bq}) results in $B\sim O(E_{\rm F}^2)$ for
$E_{\rm F}\rightarrow 0$ in this case --- a much stronger
enhancement of the q-factor than for the impurity on top of carbon
or at a bridge site.

The origin of this effect can be either understood in terms of the
$C_6$ symmetry as discussed above or in terms of destructive
quantum interference in graphene lattice:
\begin{figure}[tb]
 \includegraphics[width=0.68\linewidth]{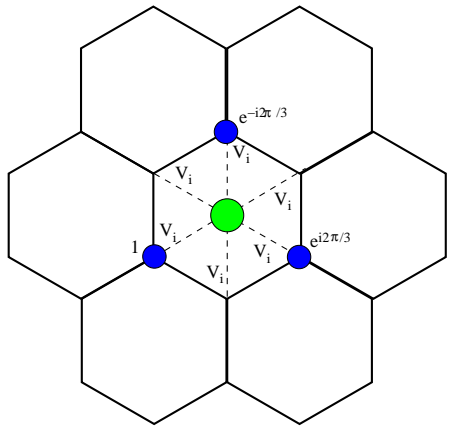}
\caption{\label{fig:cancellation} Model of an Anderson impurity in
the middle of the graphene hexagon. Electrons from neighboring can
hop onto the impurity by the hopping matrix elements $V_i$. For an
electron from the vicinity of the Brillouin zone corner $K$ phase
differences of its wave function at neighboring sites belonging to
sublattice A (blue marked atoms) are given. The sum of these phase
factors vanishes.}
\end{figure}
For each state in the vicinity of K (K') the phase of the wave
function at neighboring sites of the impurity belonging to one
sublattice circulates either clockwise or counterclockwise around
the impurity, as illustrated in Fig. \ref{fig:cancellation}. The
phase of the wave function of atoms in sublattice A having one
common nearest neighbor in sublattice B circulates around this
atom in sublattice B leading to $\xi(k)\rightarrow 0$ for
$k\rightarrow K$ or $k\rightarrow K'$. This cancellation can also
be viewed as a result of Berry phase associated with  the Dirac
points in pristine graphene. The relation between vanishing of
$\xi(k)$ and topological properties of honeycomb lattice was
discussed in Ref. \onlinecite{manes-2007}. Thus, the imaginary
part $Im G(E)$ vanishes linearly at the conical point which leads,
taking into account analytical properties of $G(E)$ to logarithmic
divergence of the $q$-factor.

In conclusion, we have addressed the Kondo effect in graphene for
a realistic d-electron case of Co atom. The crystal field of
graphene honeycomb lattice splits the d-orbitals into two
doublets, $E_{1,2}$ and one singlet $A_1$ states. We have
performed the first principles calculations and found that $E_1$
doublet states are responsible for the Kondo effect and for
unusual Fano q-factors seen in the experiments
\cite{Manoharan_Co_Graphene}. For the impurity placed in the
middle of hexagon we have found that the same destructive
interference that lead to linearly vanishing DOS and to Berry
phase is responsible for the anomalously large $q$-factors in Fano
resonance analysis. We thus conclude that nontrivial properties of
the Kondo effect in graphene are related to masless Dirac fermion
spectrum. The quantum interference effects, like the Fano effect
considered here, are extremely sensitive to atomistic details such
es specific impurity positions, which can be checked
experimentally. Upon completion  of this work we learned about the
recent preprint by H.-B. Zhuang \textit{et. al.}
http://arxiv.org/abs/0905.4548 that addresses related questions
for the case of a s-wave magnetic impurity.

\textit{Acknowledgements}.

 We are grateful to D. Arovas, A.Rosch and J. von
Delft for useful discussions. This work was supported by Stichting voor Fundamenteel
Onderzoek der Materie (FOM), the Netherlands, aby SFB-668(A3),
Germany.  Work at Los Alamos was supported by US DOE through BES and LDRD funds  under the auspices of NNSA of the U.S. Department of Energy  under contract No DE-AC52-06NA25396. The work  at  Stanford University was supported  by the Department of Energy, Office of Basic Energy Sciences, Division of Materials Sciences and Engineering, under contract DE-AC02-76SF00515.

\bibliography{graphene_s}

\begin{thebibliography}{25}
\expandafter\ifx\csname natexlab\endcsname\relax\def\natexlab#1{#1}\fi
\expandafter\ifx\csname bibnamefont\endcsname\relax
  \def\bibnamefont#1{#1}\fi
\expandafter\ifx\csname bibfnamefont\endcsname\relax
  \def\bibfnamefont#1{#1}\fi
\expandafter\ifx\csname citenamefont\endcsname\relax
  \def\citenamefont#1{#1}\fi
\expandafter\ifx\csname url\endcsname\relax
  \def\url#1{\texttt{#1}}\fi
\expandafter\ifx\csname urlprefix\endcsname\relax\def\urlprefix{URL }\fi
\providecommand{\bibinfo}[2]{#2}
\providecommand{\eprint}[2][]{\url{#2}}

\bibitem[{\citenamefont{Hewson}(1993)}]{HewsonBook}
\bibinfo{author}{\bibfnamefont{A.~C.} \bibnamefont{Hewson}},
  \emph{\bibinfo{title}{The Kondo problem to heavy fermions}}
  (\bibinfo{publisher}{Cambridge University Press}, \bibinfo{year}{1993}).

\bibitem[{\citenamefont{Madhavan et~al.}(1998)\citenamefont{Madhavan, Chen,
  Jamneala, Crommie, and Wingreen}}]{Madhavan98}
\bibinfo{author}{\bibfnamefont{V.}~\bibnamefont{Madhavan}},
  \bibinfo{author}{\bibfnamefont{W.}~\bibnamefont{Chen}},
  \bibinfo{author}{\bibfnamefont{T.}~\bibnamefont{Jamneala}},
  \bibinfo{author}{\bibfnamefont{M.~F.} \bibnamefont{Crommie}},
  \bibnamefont{and} \bibinfo{author}{\bibfnamefont{N.~S.}
  \bibnamefont{Wingreen}}, \bibinfo{journal}{Science}
  \textbf{\bibinfo{volume}{280}}, \bibinfo{pages}{567} (\bibinfo{year}{1998}),
  \eprint{http://www.sciencemag.org/cgi/reprint/280/5363/567.pdf},
  \urlprefix\url{http://www.sciencemag.org/cgi/content/abstract/280/5363/567}.

\bibitem[{\citenamefont{Knorr et~al.}(2002)\citenamefont{Knorr, Schneider,
  Diekh\"oner, Wahl, and Kern}}]{Knorr02}
\bibinfo{author}{\bibfnamefont{N.}~\bibnamefont{Knorr}},
  \bibinfo{author}{\bibfnamefont{M.~A.} \bibnamefont{Schneider}},
  \bibinfo{author}{\bibfnamefont{L.}~\bibnamefont{Diekh\"oner}},
  \bibinfo{author}{\bibfnamefont{P.}~\bibnamefont{Wahl}}, \bibnamefont{and}
  \bibinfo{author}{\bibfnamefont{K.}~\bibnamefont{Kern}},
  \bibinfo{journal}{Phys. Rev. Lett.} \textbf{\bibinfo{volume}{88}},
  \bibinfo{pages}{096804} (\bibinfo{year}{2002}).

\bibitem[{\citenamefont{Limot and Berndt}(2004)}]{LimotCM}
\bibinfo{author}{\bibfnamefont{L.}~\bibnamefont{Limot}} \bibnamefont{and}
  \bibinfo{author}{\bibfnamefont{R.}~\bibnamefont{Berndt}},
  \bibinfo{journal}{Appl. Surf. Science} \textbf{\bibinfo{volume}{237}},
  \bibinfo{pages}{572} (\bibinfo{year}{2004}).

\bibitem[{\citenamefont{N\'{e}el et~al.}(2008)\citenamefont{N\'{e}el,
  Kr\"{o}ger, Berndt, Wehling, Lichtenstein, and Katsnelson}}]{Neel}
\bibinfo{author}{\bibfnamefont{N.}~\bibnamefont{N\'{e}el}},
  \bibinfo{author}{\bibfnamefont{J.}~\bibnamefont{Kr\"{o}ger}},
  \bibinfo{author}{\bibfnamefont{R.}~\bibnamefont{Berndt}},
  \bibinfo{author}{\bibfnamefont{T.~O.} \bibnamefont{Wehling}},
  \bibinfo{author}{\bibfnamefont{A.~I.} \bibnamefont{Lichtenstein}},
  \bibnamefont{and} \bibinfo{author}{\bibfnamefont{M.~I.}
  \bibnamefont{Katsnelson}}, \bibinfo{journal}{Phys. Rev. Lett.}
  \textbf{\bibinfo{volume}{101}}, \bibinfo{eid}{266803}
  (pages~\bibinfo{numpages}{4}) (\bibinfo{year}{2008}),
  \urlprefix\url{http://link.aps.org/abstract/PRL/v101/e266803}.

\bibitem[{\citenamefont{Novoselov et~al.}(2004)\citenamefont{Novoselov, Geim,
  Morozov, Jiang, Zhang, Dubonos, Grigorieva, and
  Firsov}}]{Novoselov_science2004}
\bibinfo{author}{\bibfnamefont{K.~S.} \bibnamefont{Novoselov}},
  \bibinfo{author}{\bibfnamefont{A.~K.} \bibnamefont{Geim}},
  \bibinfo{author}{\bibfnamefont{S.~V.} \bibnamefont{Morozov}},
  \bibinfo{author}{\bibfnamefont{D.}~\bibnamefont{Jiang}},
  \bibinfo{author}{\bibfnamefont{Y.}~\bibnamefont{Zhang}},
  \bibinfo{author}{\bibfnamefont{S.~V.} \bibnamefont{Dubonos}},
  \bibinfo{author}{\bibfnamefont{I.~V.} \bibnamefont{Grigorieva}},
  \bibnamefont{and} \bibinfo{author}{\bibfnamefont{A.~A.}
  \bibnamefont{Firsov}}, \bibinfo{journal}{Science}
  \textbf{\bibinfo{volume}{306}}, \bibinfo{pages}{666} (\bibinfo{year}{2004}).

\bibitem[{\citenamefont{Novoselov et~al.}(2005)\citenamefont{Novoselov, Geim,
  Morozov, Jiang, Katsnelson, Grigorieva, Dubonos, and Firsov}}]{Geim2005}
\bibinfo{author}{\bibfnamefont{K.~S.} \bibnamefont{Novoselov}},
  \bibinfo{author}{\bibfnamefont{A.~K.} \bibnamefont{Geim}},
  \bibinfo{author}{\bibfnamefont{S.~V.} \bibnamefont{Morozov}},
  \bibinfo{author}{\bibfnamefont{D.}~\bibnamefont{Jiang}},
  \bibinfo{author}{\bibfnamefont{M.~I.} \bibnamefont{Katsnelson}},
  \bibinfo{author}{\bibfnamefont{I.~V.} \bibnamefont{Grigorieva}},
  \bibinfo{author}{\bibfnamefont{S.~V.} \bibnamefont{Dubonos}},
  \bibnamefont{and} \bibinfo{author}{\bibfnamefont{A.~A.}
  \bibnamefont{Firsov}}, \bibinfo{journal}{Nature}
  \textbf{\bibinfo{volume}{438}}, \bibinfo{pages}{197} (\bibinfo{year}{2005}).

\bibitem[{\citenamefont{Zhang et~al.}(2005)\citenamefont{Zhang, Tan, Stormer,
  and Kim}}]{Zhang2005}
\bibinfo{author}{\bibfnamefont{Y.}~\bibnamefont{Zhang}},
  \bibinfo{author}{\bibfnamefont{Y.-W.} \bibnamefont{Tan}},
  \bibinfo{author}{\bibfnamefont{H.~L.} \bibnamefont{Stormer}},
  \bibnamefont{and} \bibinfo{author}{\bibfnamefont{P.}~\bibnamefont{Kim}},
  \bibinfo{journal}{Nature} \textbf{\bibinfo{volume}{438}},
  \bibinfo{pages}{201} (\bibinfo{year}{2005}).

\bibitem[{\citenamefont{Cassanello and Fradkin}(1996)}]{fradkin-1996}
\bibinfo{author}{\bibfnamefont{C.~R.} \bibnamefont{Cassanello}}
  \bibnamefont{and} \bibinfo{author}{\bibfnamefont{E.}~\bibnamefont{Fradkin}},
  \bibinfo{journal}{Phys. Rev. B} \textbf{\bibinfo{volume}{53}},
  \bibinfo{pages}{15079} (\bibinfo{year}{1996}).

\bibitem[{\citenamefont{Sengupta and Baskaran}(2008)}]{Sengupta07}
\bibinfo{author}{\bibfnamefont{K.}~\bibnamefont{Sengupta}} \bibnamefont{and}
  \bibinfo{author}{\bibfnamefont{G.}~\bibnamefont{Baskaran}},
  \bibinfo{journal}{Physical Review B (Condensed Matter and Materials Physics)}
  \textbf{\bibinfo{volume}{77}}, \bibinfo{eid}{045417}
  (pages~\bibinfo{numpages}{5}) (\bibinfo{year}{2008}),
  \urlprefix\url{http://link.aps.org/abstract/PRB/v77/e045417}.

\bibitem[{\citenamefont{P.~S.~Cornaglia and Balseiro}(2009)}]{balseiro-2009}
\bibinfo{author}{\bibfnamefont{G.~U.} \bibnamefont{P.~S.~Cornaglia}}
  \bibnamefont{and} \bibinfo{author}{\bibfnamefont{C.~A.}
  \bibnamefont{Balseiro}}, \bibinfo{journal}{Phys. Rev. Lett.}
  \textbf{\bibinfo{volume}{102}}, \bibinfo{pages}{046801}
  (\bibinfo{year}{2009}).

\bibitem[{\citenamefont{Balatsky et~al.}(2006)\citenamefont{Balatsky, Vekhter,
  and Zhu}}]{RMP_Balatsky}
\bibinfo{author}{\bibfnamefont{A.~V.} \bibnamefont{Balatsky}},
  \bibinfo{author}{\bibfnamefont{I.}~\bibnamefont{Vekhter}}, \bibnamefont{and}
  \bibinfo{author}{\bibfnamefont{J.-X.} \bibnamefont{Zhu}},
  \bibinfo{journal}{Rev. Mod. Phys.} \textbf{\bibinfo{volume}{78}},
  \bibinfo{eid}{373} (\bibinfo{year}{2006}).

\bibitem[{\citenamefont{Schedin et~al.}(2007)\citenamefont{Schedin, Geim,
  Morozov, Hill, Blake, Katsnelson, and Novoselov}}]{SchedinGassensors}
\bibinfo{author}{\bibfnamefont{F.}~\bibnamefont{Schedin}},
  \bibinfo{author}{\bibfnamefont{A.~K.} \bibnamefont{Geim}},
  \bibinfo{author}{\bibfnamefont{S.~V.} \bibnamefont{Morozov}},
  \bibinfo{author}{\bibfnamefont{E.~W.} \bibnamefont{Hill}},
  \bibinfo{author}{\bibfnamefont{P.}~\bibnamefont{Blake}},
  \bibinfo{author}{\bibfnamefont{M.~I.} \bibnamefont{Katsnelson}},
  \bibnamefont{and} \bibinfo{author}{\bibfnamefont{K.~S.}
  \bibnamefont{Novoselov}}, \bibinfo{journal}{Nat. Mater.}
  \textbf{\bibinfo{volume}{6}}, \bibinfo{pages}{652} (\bibinfo{year}{2007}).

\bibitem[{\citenamefont{Wehling et~al.}(2008)\citenamefont{Wehling, Novoselov,
  Morozov, Vdovin, Katsnelson, Geim, and Lichtenstein}}]{NanoLettAds}
\bibinfo{author}{\bibfnamefont{T.~O.} \bibnamefont{Wehling}},
  \bibinfo{author}{\bibfnamefont{K.~S.} \bibnamefont{Novoselov}},
  \bibinfo{author}{\bibfnamefont{S.~V.} \bibnamefont{Morozov}},
  \bibinfo{author}{\bibfnamefont{E.~E.} \bibnamefont{Vdovin}},
  \bibinfo{author}{\bibfnamefont{M.~I.} \bibnamefont{Katsnelson}},
  \bibinfo{author}{\bibfnamefont{A.~K.} \bibnamefont{Geim}}, \bibnamefont{and}
  \bibinfo{author}{\bibfnamefont{A.~I.} \bibnamefont{Lichtenstein}},
  \bibinfo{journal}{Nano Lett.} \textbf{\bibinfo{volume}{8}},
  \bibinfo{pages}{173} (\bibinfo{year}{2008}).

\bibitem[{\citenamefont{Manoharan et~al.}()}]{Manoharan_Co_Graphene}
\bibinfo{author}{\bibfnamefont{H.}~\bibnamefont{Manoharan}}
  \bibnamefont{et~al.}, \bibinfo{note}{{to be published.}}

\bibitem[{\citenamefont{Wallace}(1947)}]{Wallace-1947}
\bibinfo{author}{\bibfnamefont{P.~R.} \bibnamefont{Wallace}},
  \bibinfo{journal}{Phys. Rev.} \textbf{\bibinfo{volume}{71}},
  \bibinfo{pages}{622} (\bibinfo{year}{1947}).

\bibitem[{\citenamefont{Semenoff}(1984)}]{Semenoff-1984}
\bibinfo{author}{\bibfnamefont{G.~W.} \bibnamefont{Semenoff}},
  \bibinfo{journal}{Phys. Rev. Lett.} \textbf{\bibinfo{volume}{53}},
  \bibinfo{pages}{2449} (\bibinfo{year}{1984}).

\bibitem[{\citenamefont{Madhavan et~al.}(2001)\citenamefont{Madhavan, Chen,
  Jamneala, Crommie, and Wingreen}}]{Madhavan01}
\bibinfo{author}{\bibfnamefont{V.}~\bibnamefont{Madhavan}},
  \bibinfo{author}{\bibfnamefont{W.}~\bibnamefont{Chen}},
  \bibinfo{author}{\bibfnamefont{T.}~\bibnamefont{Jamneala}},
  \bibinfo{author}{\bibfnamefont{M.~F.} \bibnamefont{Crommie}},
  \bibnamefont{and} \bibinfo{author}{\bibfnamefont{N.~S.}
  \bibnamefont{Wingreen}}, \bibinfo{journal}{Phys. Rev. B}
  \textbf{\bibinfo{volume}{64}}, \bibinfo{pages}{165412}
  (\bibinfo{year}{2001}).

\bibitem[{\citenamefont{Wehling et~al.}(2007)\citenamefont{Wehling, Balatsky,
  Katsnelson, Lichtenstein, Scharnberg, and Wiesendanger}}]{imp_loc-2007-}
\bibinfo{author}{\bibfnamefont{T.~O.} \bibnamefont{Wehling}},
  \bibinfo{author}{\bibfnamefont{A.~V.} \bibnamefont{Balatsky}},
  \bibinfo{author}{\bibfnamefont{M.~I.} \bibnamefont{Katsnelson}},
  \bibinfo{author}{\bibfnamefont{A.~I.} \bibnamefont{Lichtenstein}},
  \bibinfo{author}{\bibfnamefont{K.}~\bibnamefont{Scharnberg}},
  \bibnamefont{and}
  \bibinfo{author}{\bibfnamefont{R.}~\bibnamefont{Wiesendanger}},
  \bibinfo{journal}{Phys. Rev. B} \textbf{\bibinfo{volume}{75}},
  \bibinfo{pages}{125425} (\bibinfo{year}{2007}).

\bibitem[{\citenamefont{Perdew et~al.}(1992)}]{Perdew:PW91}
\bibinfo{author}{\bibfnamefont{J.~P.} \bibnamefont{Perdew}}
  \bibnamefont{et~al.}, \bibinfo{journal}{Phys. Rev. B}
  \textbf{\bibinfo{volume}{46}}, \bibinfo{pages}{6671} (\bibinfo{year}{1992}).

\bibitem[{\citenamefont{Kresse and Hafner}(1994)}]{Kresse:PP_VASP}
\bibinfo{author}{\bibfnamefont{G.}~\bibnamefont{Kresse}} \bibnamefont{and}
  \bibinfo{author}{\bibfnamefont{J.}~\bibnamefont{Hafner}},
  \bibinfo{journal}{J. Phys.: Condes. Matter} \textbf{\bibinfo{volume}{6}},
  \bibinfo{pages}{8245} (\bibinfo{year}{1994}).

\bibitem[{\citenamefont{Bl\"ochl}(1994)}]{Bloechl:PAW1994}
\bibinfo{author}{\bibfnamefont{P.~E.} \bibnamefont{Bl\"ochl}},
  \bibinfo{journal}{Phys. Rev. B} \textbf{\bibinfo{volume}{50}},
  \bibinfo{pages}{17953} (\bibinfo{year}{1994}).

\bibitem[{\citenamefont{Kresse and Joubert}(1999)}]{Kresse:PAW_VASP}
\bibinfo{author}{\bibfnamefont{G.}~\bibnamefont{Kresse}} \bibnamefont{and}
  \bibinfo{author}{\bibfnamefont{D.}~\bibnamefont{Joubert}},
  \bibinfo{journal}{Phys. Rev. B} \textbf{\bibinfo{volume}{59}},
  \bibinfo{pages}{1758} (\bibinfo{year}{1999}).

\bibitem[{\citenamefont{Amadon et~al.}(2008)\citenamefont{Amadon, Lechermann,
  Georges, Jollet, Wehling, and Lichtenstein}}]{PAW-DMFT}
\bibinfo{author}{\bibfnamefont{B.}~\bibnamefont{Amadon}},
  \bibinfo{author}{\bibfnamefont{F.}~\bibnamefont{Lechermann}},
  \bibinfo{author}{\bibfnamefont{A.}~\bibnamefont{Georges}},
  \bibinfo{author}{\bibfnamefont{F.}~\bibnamefont{Jollet}},
  \bibinfo{author}{\bibfnamefont{T.~O.} \bibnamefont{Wehling}},
  \bibnamefont{and} \bibinfo{author}{\bibfnamefont{A.~I.}
  \bibnamefont{Lichtenstein}}, \bibinfo{journal}{Phys. Rev. B}
  \textbf{\bibinfo{volume}{77}}, \bibinfo{eid}{205112}
  (pages~\bibinfo{numpages}{13}) (\bibinfo{year}{2008}),
  \urlprefix\url{http://link.aps.org/abstract/PRB/v77/e205112}.

\bibitem[{\citenamefont{J.~L.~Mañes and Vozmediano}(2007)}]{manes-2007}
\bibinfo{author}{\bibfnamefont{F.~G.} \bibnamefont{J.~L.~Mañes}}
  \bibnamefont{and} \bibinfo{author}{\bibfnamefont{M.~A.}
  \bibnamefont{Vozmediano}}, \bibinfo{journal}{Phys. Rev. B}
  \textbf{\bibinfo{volume}{75}} (\bibinfo{year}{2007}).

\end{thebibliography}

\end{document}